\documentstyle[aps,pra,epsf]{revtex}

\begin{document}

\title{Quantum baker map on the sphere}
\vskip 0.5truecm

\author{ Prot Pako{\'n}ski$^1$, Andrzej Ostruszka$^1$, and
  Karol \.Zyczkowski$^2$\footnote{Fulbright Fellow. Permanent address:
  Instytut Fizyki im. Mariana Smoluchowskiego, Uniwersytet
  Jagiello\'nski, ul. Reymonta 4,  30-059 Krak\'ow, Poland}}

\address{
$^1$Instytut Fizyki im. Mariana Smoluchowskiego, Uniwersytet
Jagiello\'nski,
\\ ul. Reymonta 4,  30-059 Krak\'ow, Poland}
\address{
$^2$Institute for Plasma Research, University of Maryland,
 \\  College Park,  MD 20742, USA}

\date{\today}
\maketitle

\begin{abstract}
We define a class of dynamical systems on the sphere
 analogous to the baker map on the torus.
The classical maps are
characterized by dynamical entropy equal to $\ln 2$.
We construct and investigate a family of the
corresponding quantum maps. In the simplest case of the model
the system does not possess a time reversal symmetry and
the quantum map is represented by real,
orthogonal matrices of even dimension. The
semiclassical ensemble of quantum maps, obtained
by averaging over a range of matrix sizes, displays
statistical properties characteristic of circular
unitary ensemble. Time evolution
of such systems may be studied with the help of the $SU(2)$
coherent states and the generalized Husimi distribution.
\end{abstract}

\vskip 2truecm

\newpage
\section{Introduction}

The classical baker map on the torus is often used to
illustrate properties of chaotic dynamical
systems~\cite{AA68,Reichl92,Ott93}. It may be regarded
as the simplest generalization of the Bernoulli shift in two
dimensions. The baker map is measure preserving and chaotic:
it belongs to a small class of the dynamical systems, for
which the metric Kolmogorov-Sinai entropy is positive and may
be calculated analytically. This classical dynamical system
has been generalized in several ways. The generalized baker map
was used to study properties of invariant measures and fractal
dimensions~\cite{FOY83,AY84}, while snapshot attractors appearing
in the random baker map were analyzed in~\cite{RGO90,NOA96}.
Another generalization of the model, called {\sl multibaker
map}~\cite{Ga92,VTB97}, is useful for investigating the effects
of deterministic diffusion. Chaotic dynamics of a classical
map can be described as a linear evolution of densities
transformed by an associated Frobenius--Perron operator.
Much progress has recently been
achieved~\cite{Ga92,AT92,HS92,Ga93,HD94,Fo97}
in finding the spectrum and understanding some unusual properties
of the right and the left eigenstates of the F--P operator of the
classical baker map.

A first attempt to quantize the baker map on the torus is due to
Balazs and Voros~\cite{Balazs89}. Another, more symmetric version
of the quantum map was found by Saraceno ~\cite{Saraceno90}. A
general construction allowing one to quantize a class of piecewise
linear maps on the torus (including the baker map) was given by
De~Bi\`evre, Degli~Esposti, and Giachetti~\cite{Bievre96}. While
another quantization scheme was recently proposed by Rubin and
Salwen~\cite{Rubin98}. The quantum baker map on the torus become
a standard model used to pursue the concept of semiclassical
quantization of non-integrable classical systems
\cite{Ozorio91,OCT91,Saraceno92,OTS1,OTS2,La93,SV94,La95,TVS97,KH98}
emerging from the theory of Gutzwiller~\cite{Gu71,Gu90}.
In several other papers devoted to the quantum baker map one
demonstrated sensitivity of eigenstates of the system with respect to
small perturbations~\cite{Schack93},  proposed an optical realization
of this quantum map~\cite{HKO94}, and designed a scheme of realization
of the baker map by methods of {\sl quantum computing}~\cite{Schack97}.

In this work we propose a different dynamical system: the baker map
on the sphere. In fact we define an entire family of classical systems
and the corresponding quantum maps.
On one hand, the classical version of this model
is, to our best knowledge, the first dynamical system on the sphere with
explicitly computable, positive K-S entropy. On the other, analysis of
several possible versions of the analogous quantum systems may
help in clarifying various aspects of the
 quantum--classical correspondence
for non-integrable dynamical systems.

In the simplest case of the model the map defined
on the sphere is not symmetric with respect to the reversal of time.
The Floquet operator describing the time evolution
of the corresponding quantum map is
constructed out of the Wigner rotation matrices and is represented
by orthogonal matrices of even dimension.  In the general case
the evolution operators are represented by complex unitary matrices.

Each quantum model, corresponding to a given classical system, can be
generalized by introducing into the model
an arbitrary complex phase \cite{Balazs89}. One obtains than
a one parameter family of quantum maps acting in the same Hilbert
space, which may be useful for studying various aspects of level
dynamics: curvatures, correlations of velocities and anticrossings.

Phase space features and the time evolution of the quantum map can
be suitably analyzed with help of the $SU(2)$ coherent states.
A simple generalization of the quantization scheme presented
in~\cite{Balazs89} provides four different maps (with different
spectra and traces), that seem to correspond to the same classical
system. This quantization scheme allows us to obtain two versions
of the symmetric quantum baker map on the torus different from that
introduced in~\cite{Saraceno90}.

The paper is organized as follows.
The quantum version of the baker map on the sphere is introduced
in Section II. The corresponding classical system is analyzed in
Section III, while the statistical properties of the quantum map
are discussed in Section IV. Appendix~\ref{RotProp} contains the
proof of unitarity of auxiliary matrices constructed out of the
Wigner rotation matrices. Detailed formulae for four versions of
the quantum baker map on the sphere are given in
Appendix~\ref{MapSphere}, while the entire family of the classical
and quantum models is given in Appendix~\ref{GenMapS}. An analogous
construction for the baker map on the torus is shown in
Appendix~\ref{MapTorus}.

\section{Construction of the Quantum map}

To construct the quantum map
corresponding to a given classical transformation on the
sphere we use the $N$--dimensional representation of the
rotations group. Using the angular momentum operators
commutation relation
\begin{equation}
  \left[ J_i, J_j \right] = i \epsilon_{ijk} J_k
\end{equation}
we obtain $N=2j+1$ base vectors $|j,m\rangle_i$, eigenstates
of $J^2$ and $J_i$ (i-th coordinate of $J$)
\begin{equation}
  J^2|j,m\rangle_i = j(j+1)|j,m\rangle_i,\ \ \
  J_i|j,m\rangle_i = m |j,m\rangle_i,
\end{equation}
$m=-j,\ldots,j$. To work in an even dimensional Hilbert space
${\cal H}$ we consider the half--integer values of $j$.

We design the quantum  baker map $B_S$ on the sphere in such a way
that $B_S$ commutes with $J^2$ and the quantum number $j$ is preserved.
Therefore it is convenient to simplify the notation by writing
$|m\rangle_i$ instead of $|j,m\rangle_i$. The action of the baker
map has to stretch phase space along one direction ($z$-axis) and
to squeeze along the other one (say, along the $x$-axis).
The transformation between these two basis $|\Psi\rangle_x=R
|\Psi\rangle_z$ is described by the  Wigner rotation matrix $R$
\cite{Rose57} representing a rotation around $y$-axes by the angle
$\pi/2$. We use its representation in the $|m\rangle_z$ basis
\begin{equation}
  R_{m',m} = {}_z\langle m |
  e^{-i \frac{\pi}{2} J_y} | m' \rangle_z
\end{equation}
with $m,m'=-j,\dots,j$.
This matrix is real and has the following property
\begin{equation}
  R_{m',m} = (-1)^{m-m'} R_{m,m'}
    = (-1)^{m-m'} R_{-m',-m} \ .
\end{equation}
Let us split the Hilbert space ${\cal{H}}$ into two subspaces
distinguished by its projection on $| m \rangle_z$
\begin{eqnarray}
  {\cal{H}}^S & = & \left\{ |\Psi\rangle \in {\cal{H}} : {}_z\langle m |
     \Psi\rangle=0,\ m>0 \right\} \nonumber \\
  {\cal{H}}^N & = & \left\{ |\Psi\rangle \in {\cal{H}} : {}_z\langle m |
     \Psi\rangle=0,\ m<0 \right\},
\end{eqnarray}
so that ${\cal{H}}^S$ contains the states localized on the southern
hemisphere and ${\cal{H}}^N$ -- on the northern one. Projection on
$| k \rangle_x$ leads to another partition
\begin{eqnarray}
  {\cal{H}}^W & = & \left\{ |\Psi\rangle \in {\cal{H}} : {}_x\langle k |
     \Psi\rangle=0,\ k>0 \right\} \nonumber \\
  {\cal{H}}^E & = & \left\{ |\Psi\rangle \in {\cal{H}} : {}_x\langle k |
      \Psi\rangle=0,\ k<0 \right\},
\end{eqnarray}
so ${\cal{H}}={\cal{H}}^S+{\cal{H}}^N={\cal{H}}^W+{\cal{H}}^E$. The
baker map transforms each vector $|\Psi^S\rangle\in{\cal{H}}^S$
into a vector $|\Phi^W\rangle \in{\cal{H}}^W$, with a linear
stretching by a factor of two in the $z$ direction. The effect
of this stretching can be formalized by the condition that the
$N/2$ odd (even) coefficients of the state $|\Phi^W\rangle$
expanded in the basis $| m \rangle_z$ are given by the first
$N/2$ coefficients of the state $|\Psi^S\rangle$ represented
in the same basis
\begin{equation}
  {}_z\langle m |\Psi^S\rangle =
  \sqrt{2} ~  {}_z\langle 2m+j |\Phi^W\rangle
 ~ ~ {\rm{for}} ~ ~ m=-j,\dots,-1/2.
  \label{PhiEquat}
\end{equation}
The factor $\sqrt{2}$ leads to the correct normalization.

The analogue condition for the linear squeezing in $x$-direction
would give the overcomplete system of equations with no solution.
This fact is rather intuitive because it is not possible to
map the southern hemisphere into the western one with the linear
stretching in $z$-direction and the linear squeezing in $x$.
However, we require $|\Phi^W\rangle$ to be squeezed to the
western hemisphere ${\cal{H}}^W$
\begin{equation}
  {}_x\langle k |\Phi^W\rangle = 0 \
  ~ ~ {\rm{for}}  ~ ~ k=1/2,\dots,j.
  \label{PhiCond}
\end{equation}
The above condition corresponds to a nonlinear squeezing in $x$.
We want to find a linear operator $B^{SW}$ which transforms
$|\Psi^S\rangle$ into $|\Phi^W\rangle$
\begin{equation}
  | \Phi^W \rangle_z = B^{SW} |\Psi^S\rangle_z \ .
  \label{BSWDef}
\end{equation}
To rewrite this formula into the $| k \rangle_x$ basis we
multiply its both sides by the transformation matrix $R$:
$|\Phi^W\rangle_x=RB^{SW}|\Psi^S\rangle_z$. To satisfy the
condition~(\ref{PhiCond}) all elements of the lower half
of the matrix $RB^{SW}$ must vanish. Since we want $RB^{SW}$
to act only on the subspace ${\cal{H}}^S$ we set to zero the
right half of this matrix. Denoting by $M$ the only unknown
$N/2 \times N/2$ block of the matrix $RB^{SW}$ we put
equation~(\ref{BSWDef}) in the form
\begin{equation}
  {}_z\langle k | \Phi^W \rangle = \sum_{l=-j}^{j} \
  \sum_{m=-j}^{j} \ \left[ R^{-1} \right]_{k,l}
  \left[ \begin{array}{cc}  M_{l,m} & 0 \\
  0 & 0 \end{array} \right] {}_z\langle m | \Psi^S \rangle \ ,
  \label{PhiSol}
\end{equation}
Inserting above into~(\ref{PhiEquat}) we obtain the following
conditions for the matrix $M$
\begin{equation}
  \sqrt{2} \sum_{l=-j}^{-\frac{1}{2}} \ \left[ R^{-1} \right]_{2k+j,l}
    M_{l,m} = \delta_{k,m} \ \ \ \mbox{for} \ \
    k,m=-j,\ldots,-\frac{1}{2} \ .
\end{equation}
In the appendix~\ref{RotProp} we show that the matrix
$R'$ composed of half of every even (odd) column of the
matrix $\sqrt{2}\,R$ is unitary, (since it is real, it
is also orthogonal). Consequently, we take $M_{l,m}=\sqrt{2}\,
R_{l,2m+j}=:R'_{l,m}$ ($l,m=-j\ldots-1/2$) and insert it into
(\ref{PhiSol}) constructing in this way the transformation
${B}^{SW}:\,{\cal{H}}^S\rightarrow{\cal{H}}^W$. The other matrix
$R''_{l,m}:=\sqrt{2}\,R_{l,2m-j}$ (now $l,m=1/2\ldots j$) is
also unitary, which allows us to build the second transformation
${B}^{NE}:\,{\cal{H}}^N\rightarrow{\cal{H}}^E$ in the same
manner. We define the quantum baker map on the sphere as
a linear operator $B_S={B}^{SW}+{B}^{NE}$ acting on $\cal{H}$.
In the $|m\rangle_z$ basis the quantum baker map on the sphere
takes the form
\begin{equation}
  B_S = R^{-1} \left[\begin{array}{cc} R' & 0 \\ 0 & R'' \end{array}\right].
  \label{QMapS}
\end{equation}

\begin{figure}
  \hspace*{2.5cm} \epsfbox{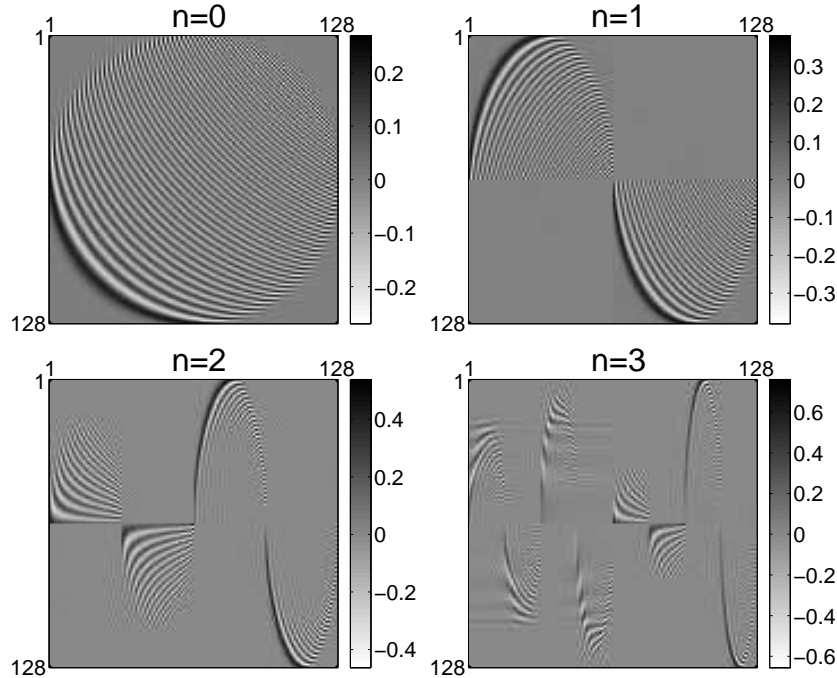} \\
  \caption{ Quantized baker map on the sphere $(B_S)^n$
    shown in the mixed representation for $N=128$ and
    $n=0,1,2$ and $3$. Gray tone scale represents values
    of real elements of $N\times N$ matrices.
    \label{FMapMix} }
\end{figure}

In the mixed representation (${}_z\langle m| B_S |m'\rangle_x$)
the baker map on the sphere consists of two diagonal blocks
$R'$ and $R''$. Figure~\ref{FMapMix} shows $(B_S)^n$ in this
representation for $n=0,1,2,3$. The first  plot obtained with
$n=0$ displays just the elements of the rotation matrix
$R_{m'm}={}_z\langle m | m'\rangle_x$. For $m$ and $m'$ fulfilling
$m^2+m'^2>j^2$ such elements are exponentially small. This
fact has a simple classical analogy. Consider a partition of
the unit ball into $N$ slices of the same width perpendicular
to the direction $z$ and another partition consisting of $N$
slices perpendicular to the direction $x$. Some side $z$-slices
(of small radii) do not overlap with some side $x$-slices.
This 'circular' structure manifests itself also in the pictures
obtained for $n=1,2$ and $3$ and contrasts with the rectangular
patterns visible in the analogous figure drawn by Saraceno
and Voros \cite{SV94} for the quantum map on the torus.

The rotation matrix $R$ is unitary, as are the matrices $R'$
and $R''$, constructed out of its elements. Hence the quantum
map $B_S$ is unitary. Writing equations~(\ref{PhiEquat}) describing
the double stretching of the vector $|\Psi^S\rangle$ we can
choose ${}_z\langle 2m+j|$ (odd column of $R$) or ${}_z\langle 2m+j+1|$
(even column) projections of the state $|\Phi^W\rangle$.
A similar ambiguity remains in the choice of the vector
$|\Phi^E\rangle$, so one may construct four different versions of
the quantum map: ${B_S}^{(00)}, {B_S}^{(01)}, {B_S}^{(10)}$ and
${B_S}^{(11)}$, as written explicitly in Appendix~\ref{MapSphere}.
In the classical limit $N\to\infty$ all four families of quantum
maps seem to tend to the same classical system.

In order to create one parameter families of quantum maps it is
possible to introduce a phase factor $\lambda$ into any of the
above four versions of the model
\begin{equation}
  B_S(\lambda) =  R^{-1}  \left[ \begin{array}{cc}
    e^{i\lambda} R'   & 0 \\ 0 &
    e^{-i\lambda} R'' \end{array} \right] .
  \label{QMapSl}
\end{equation}
This particular way of introducing the parameter $\lambda$ is
advantageous, since there is no drift of the eigenphases $\phi_i$
(i.e. the phases of complex, unimodular eigenvalues of $B_S$)
with the parameter (the mean velocity $\langle d\phi_i/d\lambda
\rangle$, averaged over individual eigenphases, vanish). An
analogous generalization of the quantum baker map on the torus
was already  proposed by Balazs and Voros~\cite{Balazs89}.
It gives a family of quantum maps corresponding to the same
classical system.

Moreover, it is possible to generalize the model by changing
the direction of squeezing, which corresponds to a different
choice of the rotation matrix $R$. As discussed in Appendix
\ref{GenMapS}, these models lead to a family of the classical 
systems, which can be parametrized by the angle $\gamma$.
The classical system corresponding to $\gamma=\pi/2$ (rotation
around the $x$ axis) possesses a generalized time reversal
symmetry, so varying the parameter $\gamma$ one can study the
effect of the time reversal symmetry breaking.

\section{Corresponding classical map}

\begin{figure}
  \Large
  \centering
  \unitlength 1.5cm
  \begin{picture}(10,2)(0,0)
    \linethickness{2pt}
    \put(2.5,0.5){\line(0,1){1}}
    \linethickness{0.3pt}
    \put(1,0.5){\framebox(3,1)[cc]{}}		
    \put(1,1){\line(1,0){3}}
    \put(1.75,1.25){\makebox(0,0)[cc]{$A_1$}}	
    \put(1.75,0.75){\makebox(0,0)[cc]{$B_1$}}
    \put(3.25,1.25){\makebox(0,0)[cc]{$A_2$}}
    \put(3.25,0.75){\makebox(0,0)[cc]{$B_2$}}
    \put(3.25,0.2){\makebox(0,0)[cc]{$\varphi$}}
    \put(1,0.3){\makebox(0,0)[cc]{$\scriptstyle 0$}}
    \put(2.5,0.3){\makebox(0,0)[cc]{$\scriptstyle \pi$}}
    \put(4,0.3){\makebox(0,0)[cc]{$\scriptstyle 2\pi$}}
    \put(0.7,1){\makebox(0,0)[cc]{$t$}}
    \put(0.7,0.5){\makebox(0,0)[cc]{$\scriptstyle -1$}}
    \put(0.8,1.5){\makebox(0,0)[cc]{$\scriptstyle 1$}}
    \put(4.3,1){\vector(1,0){1}}		
    \put(6,0.5){\framebox(3,1)[cc]{}}		
    \put(6.75,0.5){\line(0,1){1}}
    \put(7.5,0.5){\line(0,1){1}}
    \put(8.25,0.5){\line(0,1){1}}
    \put(6.375,1){\makebox(0,0)[cc]{$A'_1$}}	
    \put(7.125,1){\makebox(0,0)[cc]{$B'_2$}}
    \put(7.875,1){\makebox(0,0)[cc]{$B'_1$}}
    \put(8.625,1){\makebox(0,0)[cc]{$A'_2$}}
    \put(8.25,0.2){\makebox(0,0)[cc]{$\varphi$}} 
    \put(6,0.3){\makebox(0,0)[cc]{$\scriptstyle 0$}}
    \put(7.5,0.3){\makebox(0,0)[cc]{$\scriptstyle \pi$}}
    \put(9,0.3){\makebox(0,0)[cc]{$\scriptstyle 2\pi$}}
    \put(5.7,1){\makebox(0,0)[cc]{$t$}}
    \put(5.7,0.5){\makebox(0,0)[cc]{$\scriptstyle -1$}}
    \put(5.8,1.5){\makebox(0,0)[cc]{$\scriptstyle 1$}}
    \linethickness{0.1pt}			
    \put(4,0.5){\vector(1,0){0.3}}
    \put(1,1.5){\vector(0,1){0.3}}
    \put(9,0.5){\vector(1,0){0.3}}
    \put(6,1.5){\vector(0,1){0.3}}
    \put(1,0.75){\line(1,1){0.25}}
    \multiput(1,0.5)(0.25,0){11}{\line(1,1){0.5}}
    \put(3.75,0.5){\line(1,1){0.25}}
    \put(6.75,1.25){\line(1,1){0.25}}
    \put(6.75,1){\line(1,1){0.5}}
    \put(6.75,0.75){\line(1,1){0.75}}
    \multiput(6.75,0.5)(0.25,0){3}{\line(1,1){1}}
    \put(7.5,0.5){\line(1,1){0.75}}
    \put(7.75,0.5){\line(1,1){0.5}}
    \put(8,0.5){\line(1,1){0.25}}
  \end{picture}
  \vspace{2mm}
  \caption{ Classical baker map on the sphere in the
    coordinates ($t=\cos\theta,\varphi$). Parts $A_1$,
    $A_2$, $B_1$, $B_2$ are linearly transformed into
    $A'_1$, $A'_2$, $B'_1$, $B'_2$. \label{FCMapS} }
\end{figure}
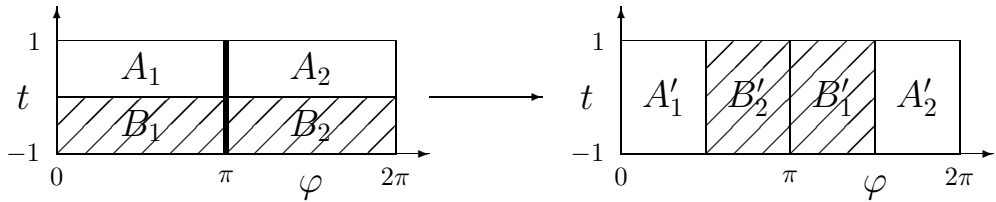

The quantum baker map on the sphere is unitary, hence the corresponding
classical map \mbox{$M:S^2\to S^2$} has to conserve the volume of the
phase space. The quantum transformation corresponds to the stretching
by a factor of two along the $z$ axis. Thus it is linear in the
variable $t=\cos\theta$. The operation of squeezing is nonlinear
in $x$, but must be linear in the angle $\phi$, which leads to the
following classical map on the sphere
\begin{equation}
  \left(t',\varphi'\right) = \left\{ \begin{array}{ll}
    \left( 2t-1, \varphi/2 \right) &
      \mbox{for $t \geq 0$ and $\varphi\leq\pi$} \\
    \left( 2t+1, \varphi/2+\pi \right) &
      \mbox{for $t < 0$ and $\varphi\leq\pi$} \\
    \left( 2t-1, \varphi/2+\pi \right) &
      \mbox{for $t \geq 0$ and $\varphi>\pi$} \\
    \left( 2t+1, \varphi/2 \right) &
      \mbox{for $t < 0$ and $\varphi>\pi$}
    \end{array} \right. ,
  \label{CMapS}
\end{equation}
It is presented in Fig.~\ref{FCMapS} in the ($t,\varphi$)
coordinates.

The classical baker map on the sphere is chaotic: the dynamical
entropy of Kolmogorov-Sinai equals to $\ln 2$. The generating
partition consists of two cells $A=A_1 \cup A_2$ and $B=B_1
\cup B_2$ as shown on Fig.~\ref{FCMapS}. The uniform measure
is invariant under the map (\ref{CMapS}) and after each
iteration all probabilities $P_{AA},P_{AB},P_{BA},P_{BB}$
of going from one cell to the other are equal. A similar
reasoning done for $n$ time steps shows that the probabilities
of all possible $2^n$ trajectories are equal to $2^{-n}$, which
leads to the same metric entropy as for the baker map on the torus.
A generalization of the model leading to a family of the classical
systems characterized by the same dynamical entropy, is given
in Appendix~\ref{GenMapS}.

Furthermore, the topological entropy (as all generalized R\'enyi
entropies \cite{Renyi70}) is equal to $\ln 2$. Therefore, the
number of points belonging periodic orbits growth with their
period $n$ as $2^n$. In particular the number $L$ of periodic
orbits of ${(B_S)}^n$ is equal to $2^n$ for $n$ odd and $2^n-2$
for $n$ even. In analogy to the work of Saraceno and
Voros~\cite{SV94} we constructed a generating function which
allows us to find the action of each trajectory, required for
the semiclassical treatment of our model\cite{PP98}.

In order to demonstrate a correspondence between the classical
and the quantum baker maps on the sphere we use the vector
coherent states~\cite{R71,A72,Perelomov86}.
Each point on the sphere, labeled by the spherical coordinates
$(\theta,\varphi)$, corresponds to the {\sl $SU(2)$ coherent state}
$| \theta, \varphi \rangle$, defined as
\begin{equation}
  | \theta,  \varphi \rangle =
   \exp\Bigr[i\theta\bigl( \sin \varphi J_x- \cos \varphi
   J_y\bigl)\Bigl] |j,j\rangle .
   \label{CSDef}
\end{equation}
Expectation values of the components of the angular momentum
operator $J$ are
\begin{equation}
  \langle \theta,\varphi|J|\theta,\varphi\rangle = j\bigl(
    \sin\theta\cos\varphi,\sin\theta\sin\varphi,\cos\theta\bigr) ,
\end{equation}
which establishes the link between the coherent state $|\theta,\varphi
\rangle$ and the vector $(\theta,\varphi)$ oriented along the direction
defined by a point on the sphere. Simple expansion of vector coherent
states in the $|m\rangle_z$ basis~\cite{He87,VS95}
\begin{equation}
  |\theta ,\varphi \rangle = \sum_{m=-j}^{m=j}
    \sin^{j-m}\left(\frac{\theta}{2}\right)
    \cos^{j+m}\left(\frac{\theta}{2}\right)
    \exp \left( i(j-m)\varphi \right)
    \left[ \left({2j \atop j-m}\right) \right]^{1/2}
    |j,m\rangle_z
\end{equation}
makes them handy to use in analytical and numerical investigations
of any quantum system corresponding to a classical map on the sphere.

\begin{figure}
  \hspace*{2.5cm} \epsfbox{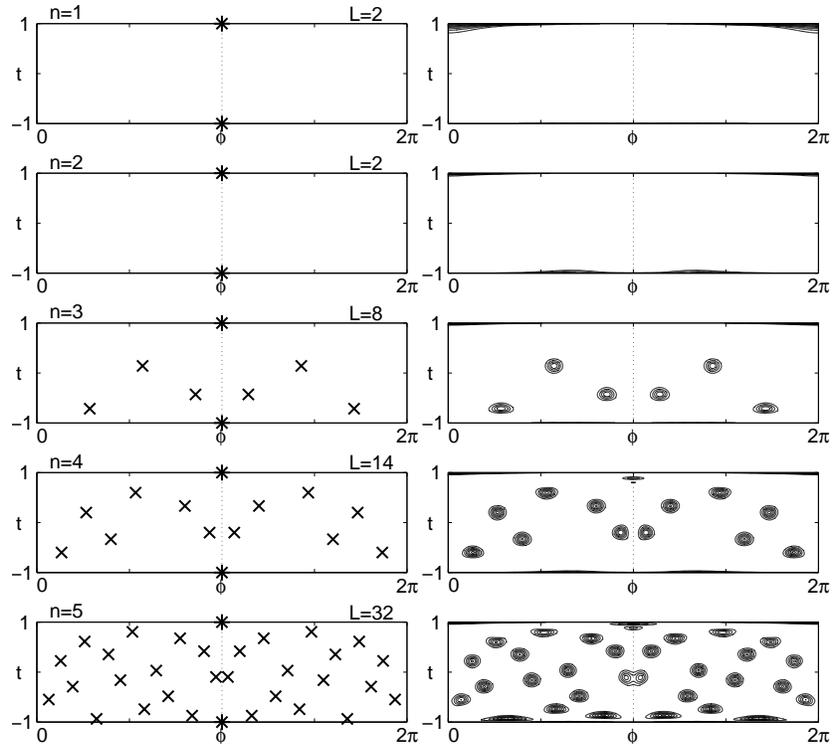} \\
  \caption{ Comparison between the classical periodical
    orbits of length $n$ (crosses in the left column)
    and squared amplitude of quantum propagator
    $|\langle\theta,\varphi|(B_S)^n|\theta,\varphi
    \rangle|^2$ (right column) for the baker map on
    the sphere. There exist $L$ periodic points of
    $(B_S)^n$ (crosses). Two of them are localized
    on the north and south poles, and are symbolically
    represented by stars arbitrarily drawn at $\phi=\pi$.
    \label{FPeriod} }
\end{figure}

Correspondence between the classical and the quantum model
is pointed out in Fig.~\ref{FPeriod}. The right column presents
contours of the function
\begin{equation}
  F_n(\theta,\varphi) = \left|\langle \theta,\varphi |
    (B_S)^n | \theta,\varphi \rangle \right|^2
\end{equation}
labeled by the number of iterations $n$. Peaks of this function
correspond to the states for which the probability of staying
after $n$ iteration of quantum map is maximal. The left column
shows periodic points of the classical map (\ref{CMapS}). The
north and south poles (represented in Fig.~\ref{FPeriod} by
horizontal lines) are fixed points of the classical map, but
the map is not continuous there. All pictures are plotted
in the ($t,\varphi$) coordinates. Quantum data are obtained
for $N=200$ and the variant ${B_S}^{(01)}$ of the model (see
appendix~\ref{MapSphere}). The analogous pictures done for
other variants of the quantum map look the same.

Analysis of eigenvectors $|v_i\rangle$ of the quantum map $B_S$
provides an additional support for the quantum--classical
correspondence. Generalized Husimi distribution of some eigenvectors,
$H_{v_i}(\theta,\varphi)=|\langle  v_i|\theta, \varphi\rangle|^2$,
reveals maxima in the vicinity of some classical periodic orbits. This
effect, called {\sl quantum scars}~\cite{He84}, was already observed
for the standard baker map on the torus\cite{Saraceno90}.

The correspondence between a given classical system $M:X\to X$ and a
family of quantum system $U_j:{\cal{H}} \to {\cal{H}}$, parametrized by
the quantum number $j$, may be quantitatively characterized by the
following condition of
{\sl regular quantization}. Let $C[(\theta,\varphi),\rho]$ denote the
circle on the sphere of radius (angle) $\rho$ centered at
$(\theta,\varphi)$. Let us define the quantity
\begin{equation}
  I_{\rho}(\theta,\varphi):= { 2j+1 \over 4 \pi}
    \int_{C[M(\theta,\varphi),\rho]}
    |\langle\theta',\varphi'|U_j|\theta,\varphi\rangle|^2
    \sin\theta' d\theta' d\varphi'
\end{equation}
measuring localization of the Husimi function of the transformed state
$U_j|\theta, \varphi\rangle$ in the $\rho$--neighborhood
of the classical image $M(\theta,\varphi)$. Due to the normalization
factor $(2j+1)/4\pi$ the integral of the Husimi function over the entire
sphere $X$ is equal to one. If for any $\rho>0$
\begin{equation}
  \lim_{N\to \infty} ~ ~ \inf_{(\theta,\varphi)} ~
    \bigl[ I_{\rho}(\theta,\varphi)\bigr] \to 1,
\end{equation}
the quantization procedure linking the classical map $M$ and the family
of quantum maps $U_j$ is called
{\sl regular} with respect to $SU(2)$ coherent states~\cite{Kief,SZ94}.
The infimum is taken over all points
$(\theta,\varphi)$ of the classical phase space, and
the size of the Hilbert space $N=2j+1$ serves as a parameter
in the family of quantum maps $U_j$.

\begin{figure}
  \hspace*{2.5cm} \epsfbox{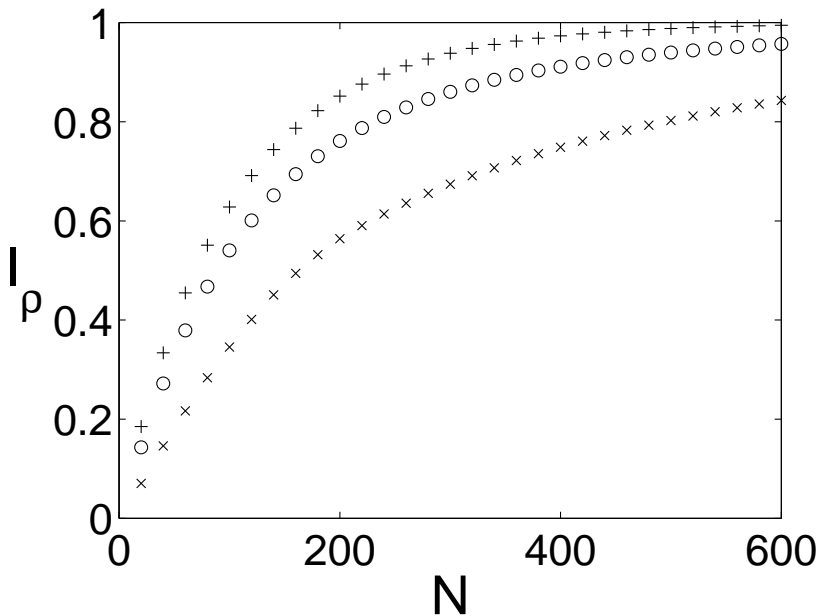} \\
  \caption{ Regular quantization condition linking the classical
    (\protect{\ref{CMapS}}) and the quantum (\protect{\ref{QMapS}})
    baker maps on the sphere. Dependence of the locally
    integrated Husimi functions $I_{\rho}$ on the size of
    the matrix $N$ for $\rho=\pi/15$. Symbol $(\circ )$
    represents the mean value averaged over $100$ points
    placed uniformly on the sphere, while the maximal and
    the minimal values are denoted by $(+)$ and $(\times)$,
    respectively. \label{FMapInt} }
\end{figure}

Not being able to prove this condition analytically, we took for $M$
and $U$ the classical (\ref{CMapS}) and the quantum (\ref{QMapS})
baker maps on the sphere, respectively, and  performed extensive
numerical tests studying the dependence of $I_{\rho}$ on $N$ and
$\rho$. Figure~\ref{FMapInt} shows how the quantity $I_{\rho}$
tends to unity in the semiclassical regime of large $N$. The mean
value of $I_{\rho}$, averaged over $100$ points on the sphere,
is close to unity at $N=600$ $(\circ)$, but the minimal value
converges much slower $(\times)$. In spite of this fact, the
above results confirm the existence of a tight relation between
the classical and  the quantum models of the baker map on the
sphere, introduced in this paper.

\section{Statistical properties of the quantum map}

Quantum maps corresponding to classically chaotic systems are
expected to display the spectral fluctuations characteristic
of random matrices~\cite{Bo91,Ha90}. Due to the lack of the time
reversal symmetry for the baker map on the sphere, unitary
matrices $B_S$ should be compared to the Dyson's circular
unitary ensemble (CUE)~\cite{Dyson62,Mehta67}. To obtain
a satisfactory statistics we accumulated data from all four
versions of the model (see appendix~\ref{RotProp}) and several
sizes of matrices.

\begin{figure}
  \hspace*{2.5cm} \epsfbox{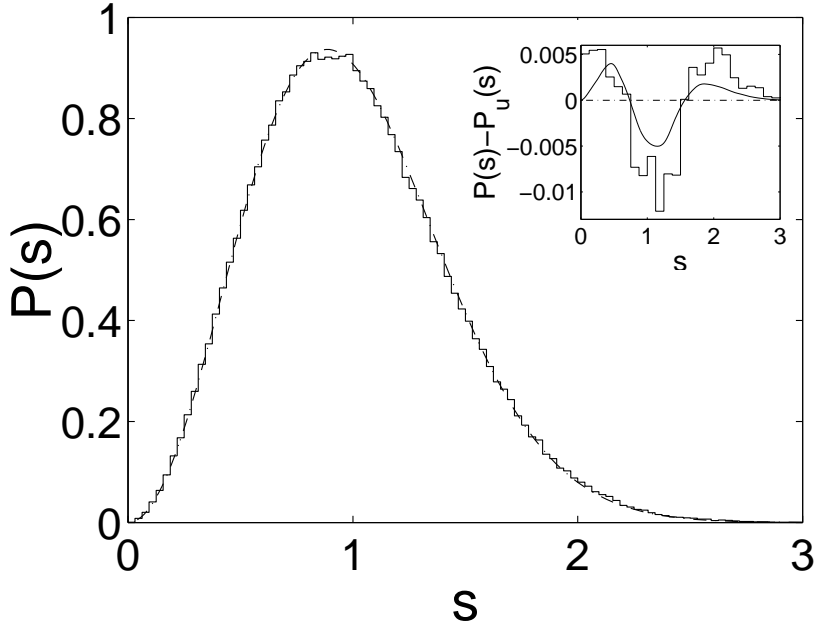} \\
  \caption{ Eigenphases spacing statistics for quantum
    baker map on the sphere. Histogram contains 79524 data
    obtained for $N=50,\ldots,800$. Dashed line represents
    the Wigner distribution $P_u(s)$. Histogram in the inset
    consists of 318096 data and exhibits the difference from
    the Wigner surmise $P(s)-P_u(s)$. The solid line
    represents exact results for random matrices
    $P_{CUE}(s)-P_u(s)$. \label{FLevSpc} }
\end{figure}

Eigenphases $\phi_l$ of
random unitary matrices are distributed uniformly in $[0,2\pi]$,
so no unfolding of the spectrum is necessary. We
numerically diagonalized matrices $B_S$,
ordered the eigenphases and computed
the rescaled spacings $s_l=(\phi_{l+1}-\phi_l)/\langle s\rangle$,
where $l=1,\dots,N-1$ and the mean spacing $\langle s \rangle$
is equal to $2\pi/N$. Figure~\ref{FLevSpc} presents the near
neighbors distribution $P(s)$ calculated from matrices $B_S$
with dimensions ranging from $N=50$ to $N=800$. The dashed line
is the Wigner distribution $P_u(s)=32\pi^{-2}s^2\exp(-4s^2/\pi^2)$
(exact for $2 \times 2$ Hermitian matrices of the Gaussian
unitary ensemble), which provides a very good approximation for the
asymptotic CUE result obtained for $N\to \infty$~\cite{Ha90}.
Collecting more data out of all four variants of the map we
observed that the precision of the Wigner surmise is not
satisfactory any more. Numerical data conform well to the exact
CUE result, implemented as a power series~\cite{DH90}, and shown
in the inset. Note that for the unitary ensemble the differences
between the Wigner surmise $P_u(s)$ and $P_{CUE}(s)$ do not
exceed $5$ parts per thousand.

\begin{figure}
  \hspace*{2.5cm} \epsfbox{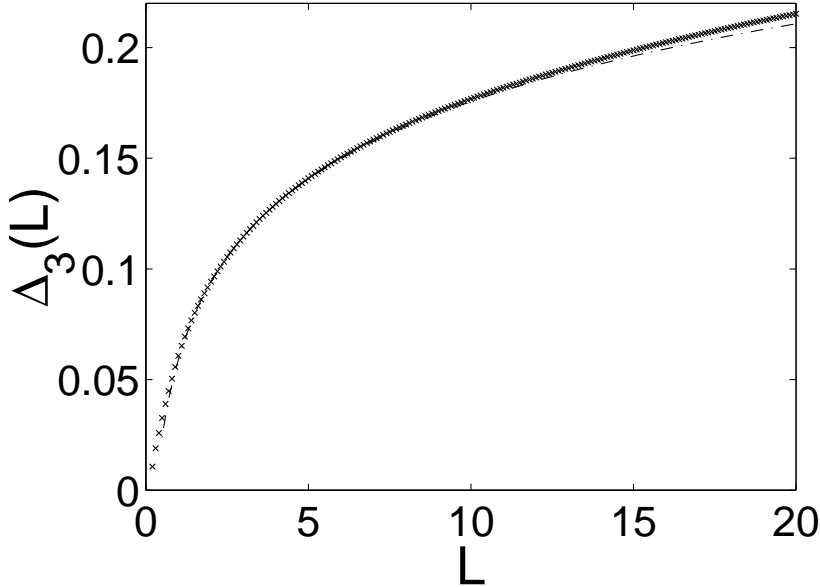} \\
  \caption{ Spectral rigidity for the quantum baker map on
    the sphere. Crosses denote	numerical data while the
    line represents the theoretical prediction for CUE.
    \label{FRigidity} }
\end{figure}

To analyze the long range correlations we computed the spectral
rigidity $\Delta_3$, introduced by Mehta and Dyson~\cite{Dyson63}.
Numerical data obtained from matrices with dimensions
$N=400\ldots 500$ are compared on Fig.~\ref{FRigidity} with CUE
results represented by a dashed line. In addition we verified
that the statistics of eigenvectors of matrices $B_S$ fit for
large $N$ to the ${\chi^2}_2$ distribution predicted for CUE
\cite{Ha90}.

Since the matrix $B_S$ is orthogonal and its spectrum consists
of two replicas of the same sequence, only the first $N/2$ spacings
out of each matrix diagonalized	 were used in the statistics.
Although the joint probability of eigenphases for random
orthogonal matrices differs from this characteristic of
CUE~\cite{Girko}, the level spacing distribution are the same
in the limit of large matrices. This fact, observed
numerically~\cite{PZK98}, can be proved in a rigorous
way~\cite{Sarnak}. Therefore it is not surprising that
the orthogonal matrices	 representing the quantum baker
map on the sphere exhibit CUE -- like  spectra.

\begin{figure}
  \hspace*{2.5cm} \epsfbox{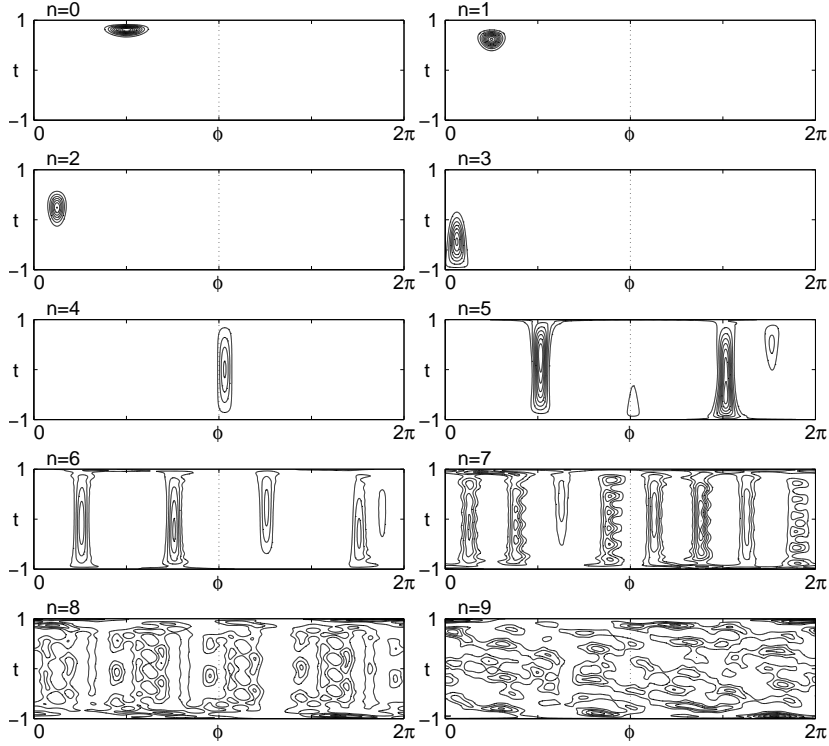} \\
  \caption{ Evolution of the coherent state localized
    at $(\frac{1}{5}\pi, \frac{1}{2}\pi)$ under action
    of the baker map $B_S$ for $N=200$. Contours of Husimi
    function are plotted in the ($t,\varphi$) coordinates.
    Note the delocalization of the wave function for
    $n>\log_2 200\approx7.6$ iteration. \label{FEvolut} }
\end{figure}

Using $SU(2)$ coherent states it is possible to visualize the
propagation of a wave packet. As before we use Husimi phase
space representation. Fig.~\ref{FEvolut} presents contours
of Husimi functions $H_{\Phi}$ with
$\Phi=(B_S)^n|\theta_0,\varphi_0\rangle$ drawn in the
($t,\varphi$) coordinates. In other words each graph
represents the subsequent iterate of the coherent state initially
localized at $(\theta_0,\varphi_0)$. After each step the wave
packet is squeezed along the $\varphi$-direction and stretched
along the $t$-direction as predict the formula~(\ref{CMapS}).
At $n=4$ the state occupies the northern and the southern
hemispheres so it is split into two part after the next
iteration. Dimension of the Hilbert space $N$ is $200$.
Observe an abrupt change in the shape of the wave packet
occurring after $7-8$ iterations. This number corresponds well
to the logarithmic time scale $n^\star=\log_2\,N\approx 7.6$,
which determines the behavior of the quantum baker map on the
torus~\cite{Saraceno90} and often emerges in several problems
of quantum chaos (see e.g.~\cite{CC96}). The same time scale
$n^\star$ was observed for all other cases investigated
($N=10,24,50,100$).

Let us also mention that a generalized version of the model,
discussed in Appendix~\ref{GenMapS}, covers also a system
enjoying the time reversal symmetry. In this case, obtained
for the parameter $\gamma$ set to $\pi/2$, the spectral
statistics of the Floquet operators coincide with the
predictions of the circular orthogonal ensemble (COE).
 
\section{Concluding remarks}

In this paper we introduced classical and quantum versions of
the baker map on the sphere. While the classical dynamics takes
place on the sphere indeed, the name {\sl quantum baker map on
the sphere} cannot be treated literally: the quantum dynamics
takes place in a finite dimensional Hilbert space $\cal{H}$,
and a link with the classical dynamics on the sphere can be
achieved with help of the $SU(2)$ coherent states. Nevertheless,
our quantum model differs, in many respects, from the
quantized baker maps of Balazs and  Voros~\cite{Balazs89} and
Saraceno~\cite{Saraceno90}. In a specific case of the model
it is represented by orthogonal
matrices, which exhibit CUE -- like statistical properties
of the spectra. This fact reflects the lack of any antiunitary
symmetry in this system. Since a time reversal case of the model,
displaying a COE - like spectra, was found, the model proposed
may be used to study the effects 
of the breaking of a generalized antiunitary symmetry.

Our construction is based on the properties of the Wigner matrix,
representing  rotation by the angle $\pi/2$. For each value of the
 classical parameter $\gamma$ we found four
different families of quantum systems, parametrized by the even
size $N$ the Hilbert space, which in the limit $N\to \infty$ seem
to correspond to the same classical system. The condition of
regular quantization with respect of $SU(2)$ coherent states
has been checked to a satisfactory precision.

In the simplest case of the model ($\gamma=0$), 
the quantum baker map on the sphere is represented by real matrices.
Thus the imaginary part of their traces vanish, in contrast to the
quantum maps on the torus which suffer a logarithmical
divergence of the imaginary
part of the trace in the semiclassical limit \cite{SV94}.

If Fourier matrices are used in our construction in place
of the Wigner rotation matrices, we obtain four versions of the
quantized baker map on the torus, some of them different than these
previously known.
We hope that the new dynamical systems will prove their
usefulness in further studies on the quantum -- classical
correspondence for chaotic systems.

\section*{Acknowledgments}
It is a pleasure to thank
F. Haake, J. Keating, M. Saraceno and J. Zakrzewski for helpful
remarks. We are indebted to an anonymous referee for several
valuable comments, which allowed us to improve and extend this work.
One of us (K\.Z) thanks Ed Ott for hospitality at the
Institute for Plasma Research, University of Maryland, where a
part of this work has been done and acknowledges the Fulbright
Fellowship. Financial support by Komitet Bada{\'n} Naukowych
under the grant no. P03B~060~13 is gratefully acknowledged.

\appendix
\section{Rotation matrix $R$ and unitarity of $R'$ }
\label{RotProp}
Components of the $N$ dimensional Wigner rotation matrix $R$,
describing rotation around $y$-axes by the angle $\pi/2$,
can be written as~\cite{Rose57}
\begin{equation}
   R_{k,m} = 2^{-j}
  \sqrt{(j+m)!(j-m)!(j+k)!(j-k)!} \ S_{k,m} \ ,
\end{equation}
where $N=2j+1$ is even and $k,m=-j,-j+1,\dots,j-1,j$.
The matrix $S_{k,m}$ reads
\begin{equation}
  S_{k,m} = \sum_{p=\max(0,k-m)}^{\min(j+k,j-m)} \
    \frac{(-1)^{p+m-k}}{p!\,(p+m-k)!\,(j-m-p)!\,(j+k-p)!} \ .
  \label{RotProp1}
\end{equation}

We want to show the following identity
\begin{equation}
  R_{-k,m} = (-1)^{j-m}	 R_{k,m} \ .
  \label{RotProp2}
\end{equation}
To this end we start computing	$R_{-k,m}$
\begin{equation}
  R_{-k,m} = 2^{-j}
    \sqrt{(j+m)!(j-m)!(j-k)!(j+k)!} \ S_{-k,m} \ .
\end{equation}
Rearranging  the sum we get
\begin{equation}
  S_{-k,m} = \sum_{p=\max(0,-k-m)}^{\min(j-k,j-m)} \
    \frac{(-1)^{p+m+k}}{p!\,(p+m+k)!\,(j-m-p)!\,(j-k-p)!} \ .
  \label{RotProp3}
\end{equation}
In~(\ref{RotProp3}) we replace $p$ by $q=j-m-p$ (so $p=j-m-q$)
and determine the new limits of summation:
$q_{\min}=j-m-p_{\max}=-m-\min(-k,-m)=\max(0,k-m)$,
while $q_{\max}=j-m-p_{\min}=j-m+\min(0,k+m)=\min(j+k,j-m)$.
Now we rewrite the sum (\ref{RotProp3})
\begin{equation}
  S_{-k,m} = \sum_{q=\max(0,k-m)}^{\min(j+k,j-m)} \
  \frac{(-1)^{j-m-q+m-k}}{(j-m-q)!\,(j+k-q)!\,q!\,(q+m-k)!} \ .
\end{equation}
Comparing the result with~(\ref{RotProp1}) we can notice that
\begin{equation}
  S_{-k,m} = (-1)^{j-m} S_{k,m} \,
\end{equation}
and the identity (\ref{RotProp2}) follows.

Unitarity of the $N/2 \times N/2$ matrix $R'$
will be deduced from the unitarity of $R$
\begin{equation}
  \sum_{k=-j}^{j} R^\dagger_{\ l,k} R_{k,m}^{} =
    \sum_{k=-j}^{j} R_{k,l} R_{k,m} = \delta_{l,m} \ .
\end{equation}
This sum may be divided into two parts
\begin{equation}
  \sum_{k=-j}^{-\frac{1}{2}} \left( R_{k,l} R_{k,m}
    + R_{-k,l} R_{-k,m} \right) = \delta_{l,m} \ .
\end{equation}
Now we use property~(\ref{RotProp2}) to reformulate the
first part of the sum
\begin{equation}
  \sum_{k=-j}^{-\frac{1}{2}} R_{k,l} R_{k,m}
    \left( 1 + (-1)^{2j-l-m} \right) = \delta_{l,m} \ .
\end{equation}
Choosing odd columns of matrix $R$ reduces to exchange
$l \rightarrow 2l+j$, and $m \rightarrow 2m+j$, so
$2j-l-m \rightarrow -2(l+m)$. Since $N$ is even, $j$ and the
indices $l$ and $m$ are half integer, so $l+m$ is an
integer number. We can rewrite the result
\begin{equation}
  2 \sum_{k=-j}^{-\frac{1}{2}}	R_{k,2l+j} R_{k,2m+j} =
    \delta_{2l+j,2m+j} = \delta_{l,m} \ .
\end{equation}
This proves that choosing upper halves of even columns of
the matrix $\sqrt{2}R$ gives an unitary $N/2 \times N/2$
matrix $R'$. Choosing odd columns or bottom halves ($k>0$)
gives other unitary matrices of the size $N/2$.

\section{Four versions of quantum baker map on the sphere}
\label{MapSphere}
As shown in appendix~\ref{RotProp} the auxiliary matrices $R'$
and $R''$ are unitary, and can be constructed out of the elements
of the rotation matrix $R$ in several ways. Let us introduce the
indices $a,b=0,1$ denoting, whether the odd or the even columns
of $R$ where used in the construction. This allows us to define
four unitary matrices
\begin{equation}
  {R'}_{k,l}^{(a)} := \sqrt{2} R_{k,2l+j+a},\ \ k,l=-j\ldots-\frac{1}{2};\ \
  {R''}_{k,l}^{(b)} := \sqrt{2} R_{k,2l-j-1+b},\ \ k,l=\frac{1}{2}\ldots j;
  \label{MapSphere1}
\end{equation}
which lead to four different variants of the
quantum baker map on the sphere for each even $N$
\begin{equation}
  {B_S}^{(ab)} = R^{-1} \left[ \begin{array}{cc}
    R'^{(a)} & 0 \\ 0 & R''^{(b)} \end{array} \right]
  \label{MapSphere2}
\end{equation}
with $a,b=0,1$. Each of matrices ${B_S}^{(ab)}$ is orthogonal and my be
generalized
into a one parameter family of unitary matrices according to formula
(\ref{QMapSl}). For concreteness we provide here some examples.
 For $N=2$ any unitary matrix is equivalent to
rotation, and formula (\ref{MapSphere2}) gives
\[ {B_S}^{(00)}={B_S}^{(10)} = \frac{1}{\sqrt{2}} \left[
   \begin{array}{cc} 1 & 1 \\ 1 & -1 \end{array} \right] \ \ \
   {\rm{and}} \ \ {B_S}^{(01)}={B_S}^{(11)} = \frac{1}{\sqrt{2}} \left[
   \begin{array}{cc} 1 & -1 \\ 1 & 1 \end{array} \right] . \]
Simplest interesting matrices appear for $N=4$
\[ {B_S}^{(00)} = \frac{1}{\sqrt{8}} \left[ \begin{array}{cccc}
   2 & 0 & 2 & 0 \\ \sqrt{3} & 1 & -\sqrt{3} & -1 \\ 0 & 2 & 0 & 2 \\
   -1 & \sqrt{3} & 1 & -\sqrt{3} \end{array} \right], \ \
   {B_S}^{(01)} = \frac{1}{\sqrt{8}} \left[ \begin{array}{cccc}
   2 & 0 & -\sqrt{3} & 1 \\ \sqrt{3} & 1 & 2 & 0 \\ 0 & 2 & -1 &
   -\sqrt{3} \\ -1 & \sqrt{3} & 0 & 2 \end{array} \right] , \]
\[ {B_S}^{(10)} = \frac{1}{\sqrt{8}} \left[ \begin{array}{cccc}
   \sqrt{3} & -1 & 2 & 0 \\ 2 & 0 & -\sqrt{3} & -1 \\ 1 & \sqrt{3}
   & 0 & 2 \\ 0 & 2 & 1 & -\sqrt{3} \end{array} \right] , \ \
   {B_S}^{(11)} = \frac{1}{\sqrt{8}} \left[ \begin{array}{cccc}
   \sqrt{3} & -1 & -\sqrt{3} & 1 \\ 2 & 0 & 2 & 0 \\ 1 & \sqrt{3}
   & -1 & -\sqrt{3} \\ 0 & 2 & 0 & 2 \end{array} \right] . \]

Observe that the first $N/2$ columns of ${B_S}^{(00)}$ and
${B_S}^{(01)}$ are equal, while the same is true for
${B_S}^{(11)}$ and ${B_S}^{(10)}$. Moreover, the matrices
${B_S}^{(00)}$ and ${B_S}^{(10)}$ share the same last $N/2$
columns; the same is true for ${B_S}^{(01)}$ and ${B_S}^{(11)}$.
These relations, valid for arbitrary even $N$, follow directly
from the definition (\ref{MapSphere2}).
To emphasize the differences between all four variants of the
quantum system let us consider the trace of each matrix. It is easy to
show that for any $N$ the matrices ${B_S}^{(10)}$ are
traceless: Tr${B_S}^{(10)}=0$. Moreover, Tr${B_S}^{(01)}=\sqrt{2}$,
while Tr${B_S}^{(00)}+$Tr${B_S}^{(11)}=\sqrt{2}$. For $N=6$ the
traceless baker map is represented by the matrix
\[ {B_S}^{(10)} = \frac{1}{8\sqrt{2}} \left[ \begin{array}{cccccc}
   3\sqrt{5} & -{\sqrt{10}} & 3 & 8 & 0 & 0 \\
   8 & 0 & 0 & - 3\sqrt{5} & - 3\sqrt{2} & 1 \\
   3\sqrt{2} & 6 & - {\sqrt{10}} & 0 & 8 & 0 \\
   0 & 8 & 0 & \sqrt{10} & -6 & - 3\sqrt{2} \\
   -1 & 3\sqrt{2} & 3\sqrt{5} & 0 & 0 & 8 \\
   0 & 0 & 8 & - 3 & \sqrt{10} & - 3 \sqrt{5}
   \end{array} \right] . \]

\section{A family of systems parametrized by a classical parameter $\gamma$}
\label{GenMapS}

Discussed quantum system may be generalized by picking for
$R$ a different rotation matrix. Let us allow for a rotation
along an arbitrary axis, which belongs to the the plane $xy$
and forms the angle $\gamma$ with the $y$-axis. The
transformation matrix from the $| m \rangle_z$ basis to the
new basis takes the form
\begin{equation}
  {R^\gamma}_{m',m} = {}_z\langle m | e^{i \gamma J_z}
    e^{-i \frac{\pi}{2} J_y} e^{-i \gamma J_z} | m' \rangle_z \ .
\end{equation}
Replacing the rotation matrix $R$ by $R^\gamma$ in the formulae
(\ref{MapSphere1}) and (\ref{MapSphere2}) we obtain a continuous
family of quantum models $B_{\gamma}^{(ab)}$ parameterized by
the angle $\gamma \in [0,2\pi)$ and corresponding to a various
directions of squeezing.
 
\begin{figure}
  \Large
  \centering
  \unitlength 1.5cm
  \begin{picture}(10,2)(0,0)
    \linethickness{2pt}
    \put(3,0.5){\line(0,1){1}}
    \linethickness{0.3pt}
    \put(1,0.5){\framebox(3,1)[cc]{}}		
    \put(1,1){\line(1,0){3}}
    \put(2,1.25){\makebox(0,0)[cc]{$A_1$}}	
    \put(2,0.75){\makebox(0,0)[cc]{$B_1$}}
    \put(3.5,1.25){\makebox(0,0)[cc]{$A_2$}}
    \put(3.5,0.75){\makebox(0,0)[cc]{$B_2$}}
    \put(3,1.67){\makebox(0,0)[cc]{$\scriptstyle \pi+2\gamma$}}
    \put(3.25,0.2){\makebox(0,0)[cc]{$\varphi$}} 
    \put(1,0.3){\makebox(0,0)[cc]{$\scriptstyle 0$}}
    \put(2.5,0.3){\makebox(0,0)[cc]{$\scriptstyle \pi$}}
    \put(4,0.3){\makebox(0,0)[cc]{$\scriptstyle 2\pi$}}
    \put(0.7,1){\makebox(0,0)[cc]{$t$}}
    \put(0.7,0.5){\makebox(0,0)[cc]{$\scriptstyle -1$}}
    \put(0.8,1.5){\makebox(0,0)[cc]{$\scriptstyle 1$}}
    \put(4.3,1){\vector(1,0){1}}		
    \put(6,0.5){\framebox(3,1)[cc]{}}		
    \put(7,0.5){\line(0,1){1}}
    \put(7.5,0.5){\line(0,1){1}}
    \put(8.5,0.5){\line(0,1){1}}
    \put(6.5,1){\makebox(0,0)[cc]{$A'_1$}}	
    \put(7.25,1){\makebox(0,0)[cc]{$B'_2$}}
    \put(8,1){\makebox(0,0)[cc]{$B'_1$}}
    \put(8.75,1){\makebox(0,0)[cc]{$A'_2$}}
    \put(7,1.67){\makebox(0,0)[cc]{$\scriptstyle \frac{1}{2}\pi+\gamma$}}
    \put(8.5,1.67){\makebox(0,0)[cc]{$\scriptstyle \frac{3}{2}\pi+\gamma$}}
    \put(8.25,0.2){\makebox(0,0)[cc]{$\varphi$}} 
    \put(6,0.3){\makebox(0,0)[cc]{$\scriptstyle 0$}}
    \put(7.5,0.3){\makebox(0,0)[cc]{$\scriptstyle \pi$}}
    \put(9,0.3){\makebox(0,0)[cc]{$\scriptstyle 2\pi$}}
    \put(5.7,1){\makebox(0,0)[cc]{$t$}}
    \put(5.7,0.5){\makebox(0,0)[cc]{$\scriptstyle -1$}}
    \put(5.8,1.5){\makebox(0,0)[cc]{$\scriptstyle 1$}}
    \linethickness{0.1pt}			
    \put(4,0.5){\vector(1,0){0.3}}
    \put(1,1.5){\vector(0,1){0.3}}
    \put(9,0.5){\vector(1,0){0.3}}
    \put(6,1.5){\vector(0,1){0.3}}
    \put(1,0.75){\line(1,1){0.25}}
    \multiput(1,0.5)(0.25,0){11}{\line(1,1){0.5}}
    \put(3.75,0.5){\line(1,1){0.25}}
    \put(7,1.25){\line(1,1){0.25}}
    \put(7,1){\line(1,1){0.5}}
    \put(7,0.75){\line(1,1){0.75}}
    \multiput(7,0.5)(0.25,0){3}{\line(1,1){1}}
    \put(7.75,0.5){\line(1,1){0.75}}
    \put(8,0.5){\line(1,1){0.5}}
    \put(8.25,0.5){\line(1,1){0.25}}
  \end{picture}
  \vspace{2mm}
  \caption{ Generalized classical baker map on the
    sphere drawn in the ($t,\varphi$) coordinates for
    $\gamma=\pi/6$. The parts $A_1$, $A_2$, $B_1$,
    $B_2$ are linearly transformed into $A'_1$,
    $A'_2$, $B'_1$, $B'_2$. \label{FGenMapS} }
\end{figure}
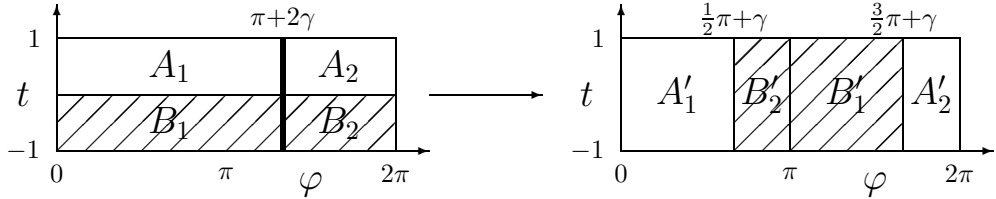

We can find the classical counterpart of the generalized
quantum model 
\begin{equation}
  \left(t',\varphi'\right) = \left\{ \begin{array}{ll}
    \left( 2t-1, \left\{ \left[(\varphi+\pi+2\gamma)
      \mbox{ mod } 2\pi\right]/2 + \frac{3}{2}\pi+\gamma \right\}
      \mbox{ mod } 2\pi \right) & \mbox{for $t \geq 0$} \\
    \left( 2t+1, \left\{ \left[(\varphi+\pi+2\gamma)
      \mbox{ mod } 2\pi\right]/2 + \frac{1}{2}\pi+\gamma \right\}
      \mbox{ mod } 2\pi \right) & \mbox{for $t < 0$}
    \end{array} \right. . \label{GenMapS1}
\end{equation}
This reduces to (\ref{CMapS}) for $\gamma=0$. Figure~\ref{FGenMapS}
shows how the sphere is mapped onto itself by the formula
(\ref{GenMapS1}). The sphere is cut along meridian
($\varphi=\pi+2\gamma$). The northern and the southern hemispheres
are stretched in the $t$-direction by the factor of
$2$ and are squeezed in the
$\varphi$-direction. The transformed blocks are respectively
placed on the east and on the west from the meridian
($\varphi=\pi/2+\gamma$).

For $\gamma=\pi/2$ the matrix $R^\gamma$ 
describes the rotation by $\pi/2$  around the $x$-axis.
 In this case the classical map is symmetric
with  respect to the reflection along the
 $x$-axis and has a generalized time
reversal symmetry. Existence of the geometric symmetry 
corresponds in the quantum case to the fact that
the Floquet operator $B_{\pi/2}^{(ab)}$ commutes 
with the reflection matrix $V_{nn'}=\delta_{n,N-1-n'}$.
Therefore one can divide its  eigenstates 
into two classes: the symmetric and the antisymmetric states. 
Analyzing the spectral statistics separately in both parity classes
we find the COE-like behavior of the spectrum. 
By varying the classical angle $\gamma$ in the vicinity of $\pi/2$
one may study the influence of the time reversal symmetry and
the geometric symmetry on the system.

\section{Four versions of quantum baker map on the torus}
\label{MapTorus}
In this appendix we obtain four variants of the
baker map on the torus and show their relation to the earlier models
\cite{Balazs89,Saraceno90}. Consider the generalized Fourier matrix
\cite{SV94} which is the transformation matrix from position $q$ to
momentum $p$ basis on the torus
\begin{equation}
  \left[{F_N}^{\chi_q\chi_p}\right]_{k,l} = \frac{1}{\sqrt{N}} \
    e^{-2\pi i(k+\chi_q)(l+\chi_p)/N}.
\end{equation}
The phases $\chi_q,\chi_p \in [0,1)$ may be treated as free
parameters, which correspond to the phases gained by the wave function
after translation by one period in the $q$ and in the $p$ direction,
respectively.
The matrix ${F_N}^{\chi_q\chi_p}$ is unitary. Assuming that
the matrix size $N$ is even,
we may apply the procedure of taking every second half
of column to obtain four auxiliary matrices of size $N/2$
\begin{equation}
  \left[{F'_{N/2}}^{\chi_q\chi_p\,(a)}\right]_{k,l} :=
  \sqrt{2}\left[{F_N}^{\chi_q\chi_p}\right]_{k,2l+a} \ ; \ \
  \left[{F''_{N/2}}^{\chi_q\chi_p\,(b)}\right]_{k,l} :=
  \sqrt{2}\left[{F_N}^{\chi_q\chi_p}\right]_{k+N/2,2l+b}
\end{equation}
where $a,b=0,1$, and the indices run $k,l=0,\ldots,N/2-1$.
Both matrices $F'^{\, (a)}$  are  unitary, since
\[ \sum_{l=0}^{N/2-1} \ \frac{2}{N} \
     e^{-2\pi i(k+\chi_q)(2l+a+\chi_p)/N}
     \,e^{2\pi i(k'+\chi_q)(2l+a+\chi_p)/N}
     = \sum_{l=0}^{N/2-1} \ \frac{2}{N} \
     e^{2\pi i(k'-k)(2l+a+\chi_p)/N} = \]
\[ = e^{2\pi i(k'-k)(a+\chi_p)/N}
     \sum_{l=0}^{N/2-1} \ \frac{1}{N/2} \
     e^{2\pi i(k'-k)l/(N/2)} = \delta_{k,k'} \ , \]
and the same holds for both matrices $F''^{\,(b)}$. Hence the four
versions of quantum baker
map on the torus take the familiar form
\begin{equation}
  {B_T}^{\chi_q\chi_p\,(ab)} := \left[{F_N}^{\chi_q\chi_p}\right]^{-1}
    \left[ \begin{array}{cc} {F'_{N/2}}^{\chi_q\chi_p\,(a)} & 0 \\
    0 & {F''_{N/2}}^{\chi_q\chi_p\,(b)} \end{array} \right].
\end{equation}

The matrix ${B_T}^{00\,(00)}$ is equivalent to the original quantum
baker map of Balazs and Voros~\cite{Balazs89}, while the symmetric map
of Saraceno~\cite{Saraceno90} is defined as follows
\begin{equation}
  B := \left[{F_N}^{1/2,1/2}\right]^{-1}
    \left[ \begin{array}{cc} {F_{N/2}}^{1/2,1/2} & 0 \\
    0 & {F_{N/2}}^{1/2,1/2} \end{array} \right].
\end{equation}
Two versions of our map ${B_T}^{1/2,1/2\,(10)}$, ${B_T}^{1/2,1/2\,(01)}$
are also symmetric in respect to reflection ($VB_TV^{-1}=B_T$ where
$V_{n,n'}=\delta_{n,N-1-n'}$), as the map of Saraceno. The real part of
their traces do not change with $N$ and is equal to $0$ and $\sqrt{2}$,
respectively. They correspond to the same classical baker map on the
torus and might be useful for further semiclassical investigation of
this model.

Statistical analysis of spectra of these two variants
${B_T}^{1/2,1/2\,(01)}$, ${B_T}^{1/2,1/2\,(10)}$ of several dimension
$N\in[50,600]$ shows that the level spacing distribution
$P(s)$ conforms to predictions of circular orthogonal ensemble with
a precision allowing one to discriminate the Wigner surmise.
The same was checked for the matrix of Saraceno. Due to symmetry of the
problem the statistical data were collected separately in
each parity class containing $N/2$ eigenvalues.

\end{document}